\pdfoutput=1
\documentclass[10pt,a4paper,altaffilletter,aps,pdftex,floatfix,longbibliography,showpacs,superscriptaddress,prl,twocolumn]{revtex4-1}

\usepackage{amsfonts,amsmath,amssymb}
\usepackage{color}
\usepackage{graphicx}
\usepackage[utf8]{inputenc}
\usepackage{hypernat}
\usepackage{mathptmx}
\usepackage[version=3]{mhchem}
\usepackage{microtype}
\usepackage{soul}
\usepackage{textcomp}
\usepackage[textsize=footnotesize,textwidth=9mm]{todonotes}
\usepackage{xspace}
\usepackage[hyperfootnotes=false]{hyperref}

\hypersetup{breaklinks,colorlinks,allcolors=black}

\newcommand{\cost}{\ensuremath{\langle\cos^2\theta_\text{2D}\rangle}\xspace}

\newcommand{\degree}{\ensuremath{^\circ}}%
\newcommand{\DIBN}{\ensuremath{\text{DIBN}}\xspace}%
\newcommand{\eg}{e.\,g.}%
\newcommand{\ie}{i.\,e.}%
\newcommand{\IYAG}{\ensuremath{\mathrm{I}_\text{YAG}}\xspace}%
\newcommand{\INoYAG}{\ensuremath{\mathrm{I}_\text{NoYAG}}\xspace}%
\newcommand{\IYAGmNoYAG}{\ensuremath{\IYAG-\INoYAG}\xspace}%
\newcommand{\Iplus}{\ensuremath{\text{I}^+}\xspace}%
\newcommand{\M}{\ensuremath{\text{M}}\xspace}%

\newlength{\figwidth}
\newlength{\figwidthsmall}
\newlength{\figwidthfull}
\newcommand{\aarhuschem}{\affiliation{Aarhus University, Department of Chemistry, 8000 Aarhus C, Denmark}}%
\newcommand{\aarhusnano}{\affiliation{Aarhus University, Interdisciplinary Nanoscience Center (iNANO), 8000 Aarhus C, Denmark}}%
\newcommand{\amolf}{\affiliation{FOM Institute AMOLF, Science Park 104, 1098 XG Amsterdam, The Netherlands}}%
\newcommand{\asu}{\affiliation{Department of Physics, Arizona State University, Tempe, AZ 85287, USA}}%
\newcommand{\cfel}{\affiliation{Center for Free-Electron-Laser Science (CFEL), DESY, Notkestrasse 85, 22607 Hamburg, Germany}}%
\newcommand{\cui}{\affiliation{Hamburg Center for Ultrafast Imaging, University of Hamburg, Luruper Chaussee 149, 22761 Hamburg, Germany}}%
\newcommand{\desy}{\affiliation{Deutsches Elektronen-Synchrotron (DESY), 22607 Hamburg, Germany}}%
\newcommand{\fhi}{\affiliation{Fritz Haber Institute of the MPG, Faradayweg 4--6, 14195 Berlin, Germany}}%
\newcommand{\ksu}{\affiliation{J.\,R.\ Macdonald Laboratory, Department of Physics, Kansas State University, Manhattan, KS 66506, USA}}%
\newcommand{\mpgasg}{\affiliation{Max Planck Advanced Study Group at CFEL, Notkestrasse 85, 22607 Hamburg, Germany}}%
\newcommand{\mpik}{\affiliation{Max Planck Institute for Nuclear Physics, 69117 Heidelberg, Germany}}%
\newcommand{\mpiep}{\affiliation{Max Planck Institute for Extraterrestrial Physics, 85741 Garching, Germany}}
\newcommand{\hll}{\affiliation{Max Planck Semiconductor Laboratory, 81739 Munich, Germany}}%
\newcommand{\lund}{\affiliation{Lund University, Department of Physics, P.\,O.\ Box 118, 22100 Lund, Sweden}}%
\newcommand{\mefo}{\affiliation{Max Planck Institute for Medical Research, 69120 Heidelberg, Germany}}%
\newcommand{\mbi}{\affiliation{Max-Born-Institute, Max Born Str.\ 2a, 12489 Berlin, Germany}}%
\newcommand{\uhh}{\affiliation{University of Hamburg, Department of Physics, Luruper Chaussee 149, 22761 Hamburg, Germany}}%
\newcommand{\unisiegen}{\affiliation{University of Siegen, Emmy-Noether Campus, Walter Flex Str.~3, 57068 Siegen, Germany}}%
\newcommand{\uppsala}{\affiliation{Uppsala University, Department of Physics and Astronomy, Box 516, 75120 Uppsala, Sweden}}%
\newcommand{\pnsensor}{\affiliation{PNSensor GmbH, 81739 Munich, Germany}}%
\newcommand{\slac}{\affiliation{SLAC National Accelerator Laboratory, Menlo Park, CA 94025, USA}}%
\newcommand{\tub}{\affiliation{Technical University of Berlin, 10623 Berlin, Germany}}%
\newcommand{\altarc}{\altaffiliation[Present address:~]{ARC Centre of Excellence for Coherent X-ray
      Science, School of Physics, The University of Melbourne, Australia}}%
\newcommand{\altbruker}{\altaffiliation[Present address:~]{Bruker AXS GmbH, Karlsruhe, Germany}}%
\newcommand{\altdesy}{\altaffiliation[Present address:~]{Deutsches Elektronen-Synchrotron (DESY),
      22607 Hamburg, Germany}}%
\newcommand{\altnijm}{\altaffiliation[Present address:~]{Institute for Molecules and Materials,
      Radboud University Nijmegen, Heijendaalseweg 135, 6525 AJ Nijmegen, The Netherlands}}%
\newcommand{\altpnsensor}{\altaffiliation[Present address:~]{PNSensor GmbH, 81739 Munich, Germany}}%
\newcommand{\altptb}{\altaffiliation[Present address:~]{Physikalisch-Technische Bundesanstalt,
      Bundesallee 100, 38116 Braunschweig, Germany}}%
\newcommand{\altpulse}{\altaffiliation[Present address:~]{Stanford PULSE Institute, SLAC National
      Accelerator Laboratory, 2575 Sand Hill Road, Menlo Park, California 94025, USA}}%
\newcommand{\altxfel}{\altaffiliation[Present address:~]{European X-ray Free Electron Laser (XFEL)
      GmbH, 22761 Hamburg, Germany}}%
\newcommand{\titletext}{X-ray diffraction from isolated and strongly aligned gas-phase molecules with a free-electron laser}%

\begin{document}
\author{Jochen Küpper}%
\email{jochen.kuepper@cfel.de}%
\homepage{http://desy.cfel.de/cid/cmi}%
\cfel\uhh\cui\fhi\mpgasg%
\author{Stephan Stern}\cfel\uhh%
\author{Lotte Holmegaard}\cfel\aarhuschem%
\author{Frank Filsinger}\altbruker\fhi\mpgasg%
\author{Arnaud Rouzée}\amolf\mbi%
\author{Artem Rudenko}\mpgasg\mpik\ksu%
\author{Per Johnsson}\lund%
\author{Andrew V. Martin}\altarc\cfel%
\author{Marcus Adolph}\tub%
\author{Andrew Aquila}\cfel%
\author{Sa\v{s}a Bajt}\cfel%
\author{Anton Barty}\cfel%
\author{Christoph Bostedt}\slac%
\author{John Bozek}\slac%
\author{Carl Caleman}\cfel\uppsala%
\author{Ryan Coffee}\slac%
\author{Nicola Coppola}\cfel%
\author{Tjark Delmas}\cfel%
\author{Sascha Epp}\mpgasg\mpik%
\author{Benjamin Erk}\altdesy\mpgasg\mpik%
\author{Lutz Foucar}\mpgasg\mefo%
\author{Tais Gorkhover}\tub%
\author{Lars Gumprecht}\cfel%
\author{Andreas Hartmann}\pnsensor%
\author{Robert Hartmann}\pnsensor%
\author{Günter Hauser}\hll\mpiep%
\author{Peter Holl}\pnsensor%
\author{Andre Hömke}\mpgasg\mpik%
\author{Nils Kimmel}\hll%
\author{Faton Krasniqi}\mpgasg\mefo%
\author{Kai-Uwe Kühnel}\mpik%
\author{Jochen Maurer}\aarhuschem%
\author{Marc Messerschmidt}\slac%
\author{Robert Moshammer}\mpik\mpgasg%
\author{Christian Reich}\pnsensor%
\author{Benedikt Rudek}\altptb\mpgasg\mpik%
\author{Robin Santra}\cfel\uhh\cui%
\author{Ilme Schlichting}\mefo\mpgasg%
\author{Carlo Schmidt}\mpgasg%
\author{Sebastian Schorb}\tub%
\author{Joachim Schulz}\altxfel\cfel%
\author{Heike Soltau}\pnsensor%
\author{John C.\ H.\ Spence}\asu%
\author{Dmitri Starodub}\altpulse\asu%
\author{Lothar Strüder}\altpnsensor\hll\unisiegen%
\author{Jan Thøgersen}\aarhuschem%
\author{Marc J.\ J.\ Vrakking}\amolf\mbi%
\author{Georg Weidenspointner}\hll\mpiep%
\author{Thomas A. White}\cfel%
\author{Cornelia Wunderer}\desy%
\author{Gerard Meijer}\altnijm\fhi%
\author{Joachim Ullrich}\altptb\mpik\mpgasg%
\author{Henrik Stapelfeldt}\aarhuschem\aarhusnano%
\author{Daniel Rolles}\mpgasg\mefo\desy%
\author{Henry N. Chapman}\cfel\uhh\cui%
\date{\today}%
\title{\titletext}
\begin{abstract}\noindent%
   We report experimental results on x-ray diffraction of quantum-state-selected and strongly
   aligned ensembles of the prototypical asymmetric rotor molecule 2,5-diiodobenzonitrile using the
   Linac Coherent Light Source. The experiments demonstrate first steps toward a new approach to
   diffractive imaging of distinct structures of individual, isolated gas-phase molecules. We
   confirm several key ingredients of single molecule diffraction experiments: the abilities to
   detect and count individual scattered x-ray photons in single shot diffraction data, to deliver
   state-selected, \eg, structural-isomer-selected, ensembles of molecules to the x-ray interaction
   volume, and to strongly align the scattering molecules. Our approach, using ultrashort x-ray
   pulses, is suitable to study ultrafast dynamics of isolated molecules.
\end{abstract}
\maketitle%
\noindent%
X-ray Free-Electron Lasers (XFELs) hold the promise to determine atomically resolved structures and
to trace structural dynamics of individual molecules and nanoparticles~\cite{Neutze:Nature406:752}.
Over the last decade, ground-breaking experiments were performed at the Free-Electron Laser at DESY
in Hamburg (FLASH)~\cite{Andruszkow:PRL85:3825, Chapman:NatPhys2:839, Barty:NatPhoton2:415,
   Jiang:PRL105.263002} and the Linac Coherent Light Source (LCLS) at the SLAC National Accelerator
Laboratory~\cite{Emma:NatPhoton4:641, Bostedt:JPB46:164003, Young:Nature466:56,
   Chapman:Nature470:73, Seibert:Nature470:78, Glownia:OE18:17620, Boll:PRA88:061402}. These
experiments already begin to provide new insights into fundamental aspects of matter, such as
hitherto unobserved structures of non-crystallizable mesoscopic objects~\cite{Barty:ARPC64:415,
   Loh:Nature486:7404, Gorkhover:PRL108:245005} or the radiation damage induced by the short and
very strong x-ray pulses~\cite{Barty:NatPhoton6:35, Young:Nature466:56}. However, the path to actual
determination of atomically resolved structures and dynamics of single molecules is still
long~\cite{Barty:ARPC64:415}. Nevertheless, related experiments on the investigation of
small-molecule structures and their dynamics utilizing molecular ensembles are within
reach~\cite{Filsinger:PCCP13:2076, Barty:ARPC64:415}.

To be able to record structural changes during ultrafast molecular processes under well-defined
conditions it was proposed~\cite{Filsinger:PCCP13:2076, Barty:ARPC64:415} to spatially separate
shapes~\cite{Helden:Science267:1483}, sizes~\cite{Trippel:PRA86:033202}, or individual
isomers~\cite{Filsinger:ACIE48:6900, Kierspel:CPL591:130, Filsinger:PRL100:133003} of complex small
molecules before delivery to the interaction point of an XFEL. The molecules should be one- or
three-dimensionally aligned or oriented in space~\cite{Stapelfeldt:RMP75:543, Spence:PRL92:198102,
   Spence:ActaCrystA61:237, Peterson:APPL92:094106, Glownia:OE18:17620, Pabst:PRA81:043425,
   Filsinger:PCCP13:2076, Hensley:PRL109:133202, Boll:PRA88:061402}. This controlled-delivery
approach would allow for the averaging of many identical patterns, similar to recent electron
diffraction experiments on aligned CF$_3$I~\cite{Hensley:PRL109:133202} or to photoelectron imaging
of 1-Ethynyl-4-fluorobenzene~\cite{Boll:PRA88:061402}. A controlled variation of the alignment
direction in space allows to tomographically build up the complete three-dimensional diffraction
volume of individual isomers. This ensemble- and pulse-averaging approach would allow working at
appropriately low fluences to circumvent detrimental electronic damage processes that have been
predicted~\cite{Lorenz:PRE86:051911, Ziaja:NJP14:115015} for the very high x-ray fluences necessary
to obtain classifiable single-molecule diffraction patterns. The forthcoming European XFEL facility
will give the opportunity to collect patterns at a rate of 27\,000 per second, which could be
sufficient to collect the necessary $10^5$--$10^8$ patterns within minutes or
hours~\cite{Barty:ARPC64:415}.

\begin{figure}
   \centering
   \includegraphics[width=\figwidth]{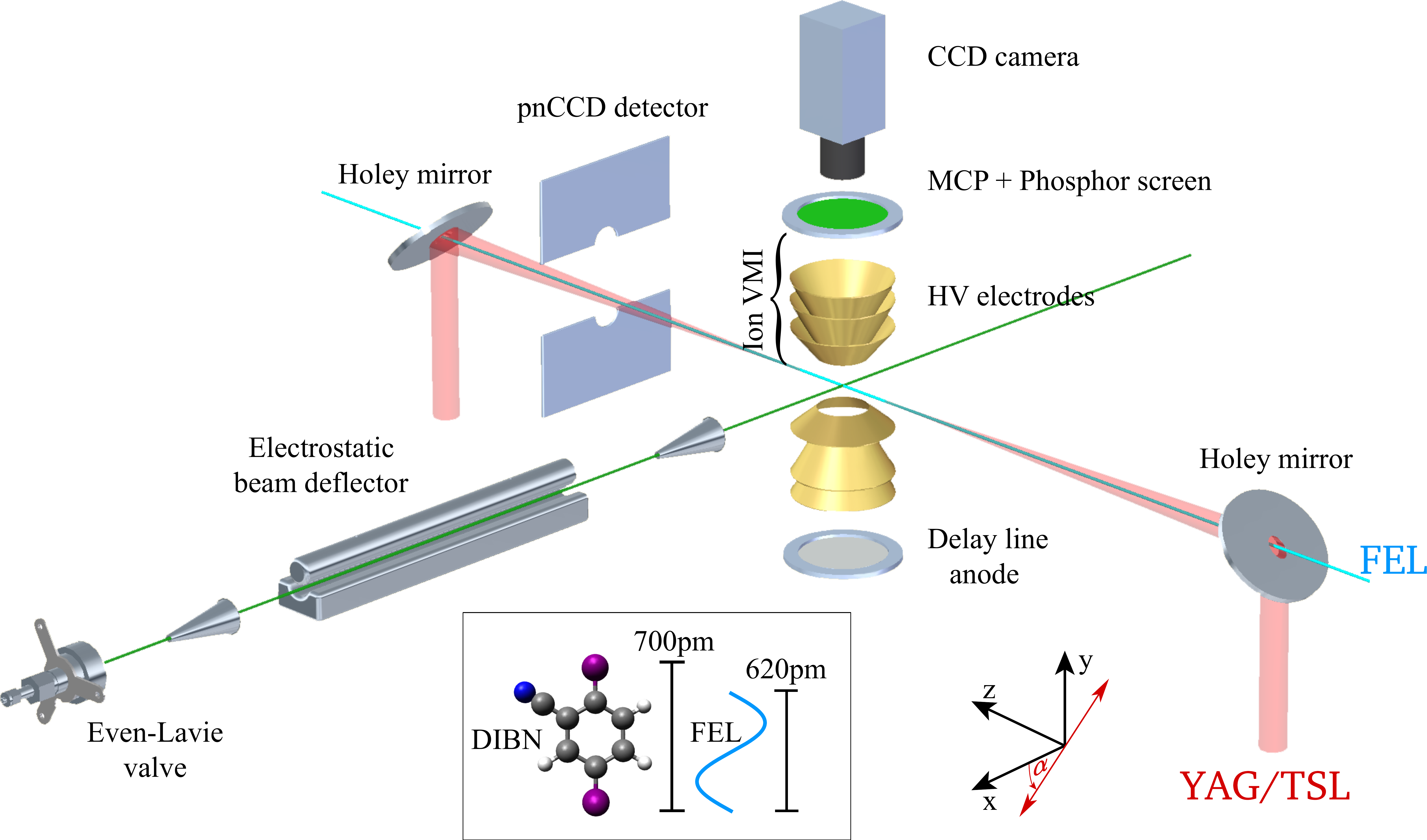}
   \caption{Schematic view of the experimental setup: from the left a supersonic beam with
      quantum-state selected molecules is delivered to the interaction point. In the center of a
      dual velocity map imaging spectrometer the molecular beam is crossed by laser beams
      copropagating from right to left. The direct laser beams go through a gap in the pnCCD
      detectors that are used to record the diffraction pattern. The upper pnCCD panel is further
      away from the beam axis than the bottom panel in order to cover a wider range of scattering
      angles. In the inset, the molecular structure of 2,5-diiodo-benzonitrile is depicted, together
      with a scale of its size, \ie, the iodine--iodine distance, and the wavelength of the x-rays.}
   \label{fig:setup}
\end{figure}

Here, we record x-ray-diffraction patterns of ensembles of identical, state-selected and strongly
aligned 2,5-diiodo-benzonitrile (\DIBN, \autoref{fig:setup}) molecules in the gas phase,
demonstrating the applicability of this controlled-delivery approach. Using 2~keV (620~pm) radiation
from the LCLS we succeeded to observe the two-center interference between the two iodine scattering
centers, separated by approximately 700~pm, in the continuous coherent diffraction pattern. The
strongly aligned samples~\cite{Holmegaard:PRL102:023001} allow to simply average the continuous
diffraction patterns from a very large number of isolated molecules~\cite{Filsinger:PCCP13:2076,
   Barty:ARPC64:415}. We restricted the angular control to one-dimensional alignment of the axis
containing the two iodine atoms, as this was the solely required control for this experiment. The
extension to three-dimensional alignment and orientation is straightforward for the cold,
state-selected samples employed~\cite{Larsen:PRL85:2470, Nevo:PCCP11:9912, Hansen:JCP139:234313}.
Moreover, we have previously demonstrated that for more complex molecules we could also exploit the
current setup to spatially separate structural isomers and sizes~\cite{Filsinger:ACIE48:6900,
   Trippel:PRA86:033202, Kierspel:CPL591:130}.

The experiment was performed at the AMO beamline at LCLS~\cite{Emma:NatPhoton4:641,
   Bostedt:JPB46:164003} using the CAMP endstation~\cite{Strueder:NIMA614:483, Foucar:CPC183:2207}
extended by a state-of-the-art molecular beam setup~\cite{Filsinger:JCP131:064309}.
\autoref{fig:setup} shows a scheme of the experimental arrangement. The setup contains multiple
devices to simultaneously detect photons, electrons, and ions~\cite{Strueder:NIMA614:483}. A pulsed
cold molecular beam is formed by expanding a few mbar of \DIBN\ in 50~bar of helium into vacuum
through an Even-Lavie valve~\cite{Even:JCP112:8068}. The molecular beam travels through an
electrostatic deflector, which disperses the molecules according to their rotational quantum states,
into the target region. There it is crossed by three pulsed laser beams: One laser beam consisting
of 12~ns (FWHM) pulses from a Nd:YAG laser (YAG, $\lambda=1064$~nm, $E_I=200$~mJ,
$\omega_0=63$~$\mu$m, $I_0\approx\!2.5\times10^{11}$~W/cm$^2$) is used to align the molecules. A
second laser beam consists of 60~fs (FWHM) pulses from a Ti:Sapphire laser (TSL, 800~nm,
$E_I=400$~$\mu$J, $\omega_0=40$~$\mu$m, $I_0\approx\!2.5\times10^{14}$~W/cm$^2$) and is used to
optimize the molecular beam and the alignment without LCLS. The third beam consists of the 100-fs
x-ray pulses (LCLS, $\lambda=620$~pm (2~keV), $E_I=4$~mJ, $\omega=30$~$\mu$m,
$I_0\approx2\cdot10^{15}$~W/cm$^2$); we estimate that 35\,\% of the generated $1.25\cdot10^{13}$
x-ray photons/pulse are transported to the experiment~\cite{Rudek:NatPhoton6:858}. All three laser
beams are copropagating, overlapped using dichroic (1064~nm and 800~nm) and holey (NIR lasers and
x-rays) mirrors before they intersect the sample and finally leave the setup through a gap in an
on-axis pnCCD camera and another holey mirror to separate the laser beams again. Time-of-flight and
velocity-map-imaging (VMI) spectrometers are installed perpendicular to the horizontal plane of the
molecular and laser beams to investigate the ion- and electron-momentum distributions resulting from
the Coulomb explosion due to absorption of one or a few x-ray photons.

\begin{figure}
   \centering
   \includegraphics[width=0.75\figwidth]{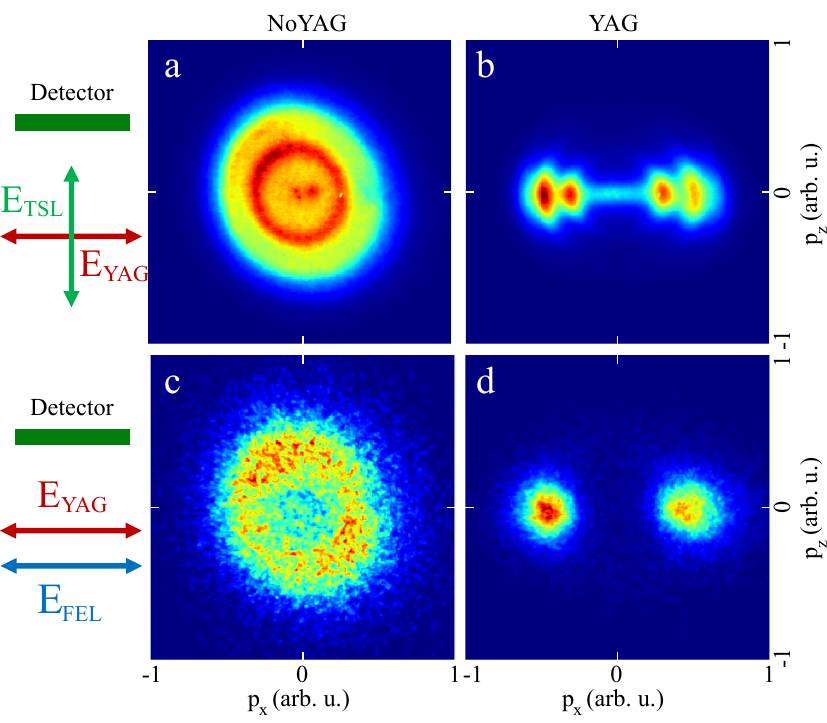}
   \caption{\Iplus\ ion images recorded with the ion-VMI detector when (a, b) the TSL or (c, d) the
      LCLS ionize and Coulomb explode the molecules. In (a, c) cylindrically symmetric distributions
      from isotropic ensembles are observed (the images are slightly distorted due to varying
      detector efficiencies). In (b, d) the horizontal alignment of the molecules, induced by the
      YAG, is clearly visible. In all measurements the YAG and the LCLS are linearly polarized
      horizontally, parallel to the detector plane, and the TSL is linearly polarized perpendicular
      to the VMI detector plane.}
   \label{fig:alignment}%
\end{figure}

We exploit Coulomb explosion imaging of DIBN induced either by strong field ionization using the TSL
pulse or through one- or two-photon ionization by the x-ray pulse to analyze the alignment of the
rotational-state-selected molecules along their I--I axis. The pertinent experimental observable is
the emission direction of the recoiling \Iplus\ ions from the Coulomb explosion, illustrated by the
2D \Iplus\ ion images in \autoref{fig:alignment}.
Without the YAG pulse the \Iplus\ images (\autoref{fig:alignment}\,a, c) were circularly symmetric
as expected for randomly aligned molecules. The circularly symmetric image obtained following
ionization with the horizontally polarized LCLS beam demonstrated that the interaction of the
far-off resonant radiation with the molecule was independent of the angle between the molecular axis
and the x-ray polarization direction: The x-rays were a practically unbiased ideal probe of the
spatial orientation of the molecules. When the YAG pulse was included the \Iplus\ ions were strongly
confined along the YAG polarization axis demonstrating tight adiabatic 1D alignment. From the
corresponding 2D momentum distribution shown in \autoref{fig:alignment}\,b and d, we extracted
$\cost=0.89$ and $\cost=0.88$ for the TSL and LCLS ionization, respectively. This degree of
alignment is in good agreement with previous measurements of adiabatic alignment of similar
molecules~\cite{Holmegaard:PRL102:023001} and stronger than previous aligment experiments of
diatomic molecules at the LCLS~\cite{Glownia:OE18:17620}. This demonstrated strong alignment of
complex molecules, even under the constraint conditions of a temporary setup at a FEL beamline. It
was made possible by the very cold molecular beam and the adiabatic alignment conditions. The
demonstrated degree of alignment fulfills the requirements to observe aligned molecule
diffraction~\cite{Spence:PRL92:198102, Filsinger:PCCP13:2076, Hensley:PRL109:133202}.

In subsequent experiments we recorded the x-ray diffraction data of these aligned samples on the
pnCCD cameras. For these experiments the polarization of the YAG laser was rotated clockwise by
$\alpha=-60\degree$. VMI data were repeatedly recorded in between diffraction experiments under the
same conditions as in \autoref{fig:alignment}. An average value for the degree of alignment in the
diffraction data of $\cost=0.84$ was derived, limited by the (changing) spatial overlap of the foci
of the YAG and the LCLS beams. The obtained x-ray diffraction patterns are shown in Fig.~S1 in the
supplementary information (SI). We have analyzed diffraction data for $\approx563\,000$ shots with
YAG and $\approx842\,000$ shots without (NoYAG), respectively, corresponding to 7~h (YAG) and 9~h
(NoYAG) measurement time with LCLS operating at 60~Hz. This data is corrected for background and
camera artifacts and individual photon hits are extracted (see SI). This results in
0.20~photons/shot which are histogrammed into the molecular diffraction pattern (Fig.~S2). By
subtracting the diffraction pattern of randomly-oriented molecules (\INoYAG) from the diffraction
pattern of aligned molecules (\IYAG), the background is cancelled. This includes the isotropic
background originating from atomic scattering of the atoms in the DIBN molecule and the helium seed
gas, as well as experimental background, \eg, scattering from apertures and rest gas.

\begin{figure}[t]
   \centering%
   \includegraphics[width=\figwidth]{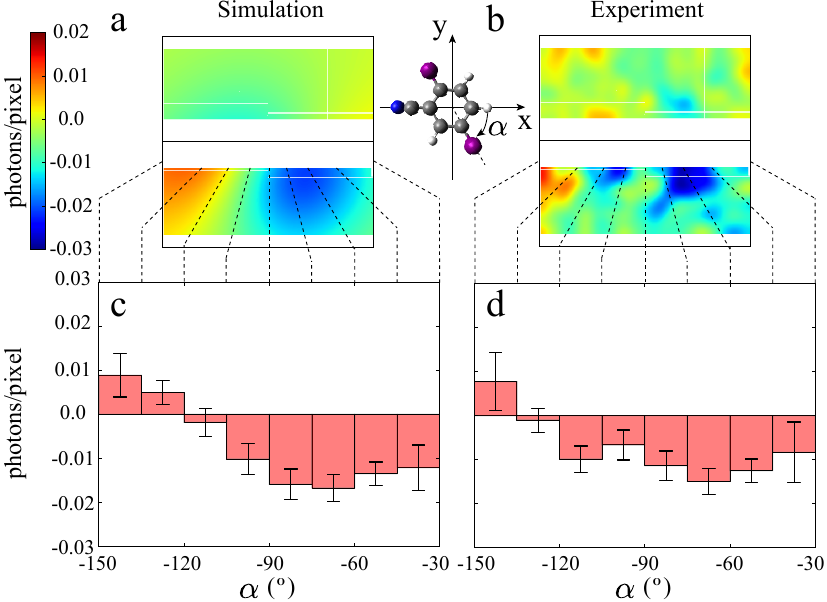}
   \caption{Diffraction-difference \IYAGmNoYAG of x-ray scattering in simulated (a) and experimental
      (b) x-ray-diffraction patterns. Histograms of the corresponding angular distributions on the
      bottom pnCCD are shown in c) and d), respectively. Error bars correspond to $1\sigma$
      statistical errors.}
   \label{fig:diffraction}%
\end{figure}
In \autoref{fig:diffraction} we present these diffraction-differences (\IYAGmNoYAG) for simulated
(\autoref{fig:diffraction}~a, c) and experimentally observed (\autoref{fig:diffraction}~b, d) x-ray
diffraction patterns. The \INoYAG data has been scaled to match the number of shots in the \IYAG
case. The anisotropy mainly originates in the scattering interference of the two (heavy) iodine
atoms. Parts of the zeroth order scattering maximum and the first minimum (along the alignment
direction $\alpha=-60\degree$) show up most prominently on the bottom pnCCD panel. The simulated
\IYAGmNoYAG image has been calculated for a molecular beam density \M~of DIBN molecules of $\M = 0.8
\cdot 10^8$~cm$^{-3}$. The error bars $\sigma$ correspond to the statistical errors from the
$\IYAG-\INoYAG$ subtraction ($\sigma = \sqrt{\IYAG+\INoYAG}$). The histograms
\autoref{fig:diffraction} (c--d) visualize the angular anisotropy which is well beyond the
statistical error in the experimentally observed image (\autoref{fig:diffraction} d), confirming the
observation of x-ray diffraction from strongly aligned samples of \DIBN.

\begin{figure}[t]
   \centering%
   \includegraphics[width=\figwidth]{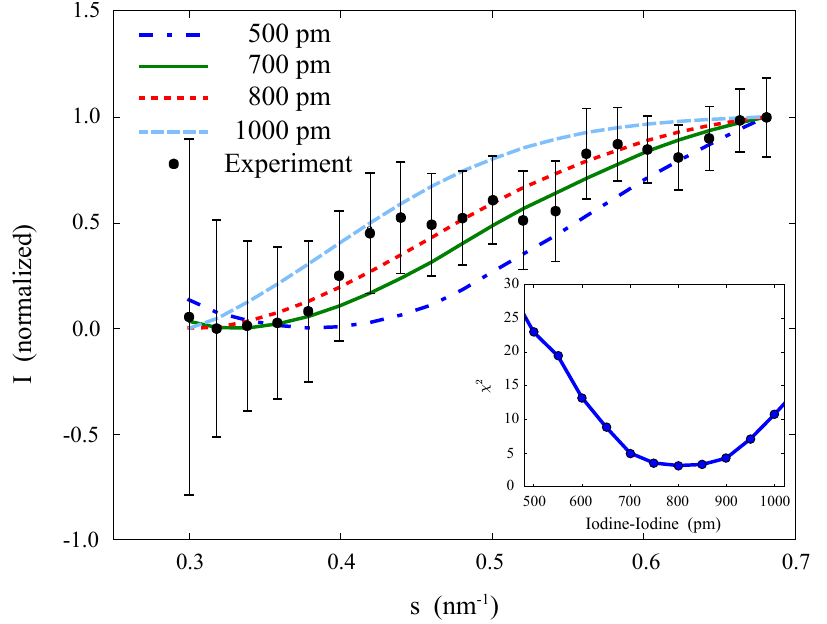}
   \caption{Comparison of experimentally obtained intensity profiles $I(s)$ along the alignment
      direction of the diffraction-difference pattern \IYAGmNoYAG with simulated profiles. The
      experimentally obtained $I(s)$ is best fitted (in terms of a $\chi^2$ test) with the model for
      an iodine-iodine distance of 800~pm (inset: test-statistic $\chi^2$ in dependence of the
      iodine-iodine distance).}
   \label{fig:analysis}%
\end{figure}

To analyze which structural information can be derived from the x-ray diffraction of isolated \DIBN
molecules, the intensity $I(s)$ in dependence of the scattering vector $s=\sin(\Theta)/\lambda$
along the alignment direction $\alpha=-60\degree$ is compared to simulated models of different
iodine-iodine distances. $\Theta$ is the scattering angle and $2\Theta$ is the angle between the
beam direction and a given detector point~\cite{Waasmaier:ActaCrystA51:416}. \emph{Ab initio}
calculations (GAMESS-US MP2/6-311G**~\cite{Schmidt:JCC14:1347}) predict a value of 700~pm for the
iodine-iodine distance. \autoref{fig:analysis} shows the experimentally obtained intensity profiles
$I(s)$, averaged over $-70\degree\le\alpha\le-50\degree$, together with simulated $I(s)$ profiles.
Each curve is normalized to be independent of the exact molecular beam density \M~of DIBN molecules,
which merely changes the contrast, \ie, the depth of the minimum. Due to the relatively long
wavelength (620~pm) compared to the known iodine-iodine distance (700~pm), the scattering extends to
large angles and the first scattering maximum from the iodine-iodine interference is not covered by
the detector in our setup. The experimentally obtained $I(s)$ is best fitted for an iodine-iodine
distance of 800~pm. \autoref{fig:analysis} shows the simulated $I(s)$ for iodine-iodine distances of
500, 700, 800, and 1000~pm. The inset of \autoref{fig:analysis} depicts the calculated
$\chi^2$-values~\cite{Nakamura:JPG37:075021:33} in dependence of the iodine-iodine distance. Due to
the experimental parameters, as mentioned above, the scattering features are large and vary only
slightly within the recorded range of $s$-values. We note that the structural features of small
molecules, like DIBN, could be determined much more accurately with data recorded at shorter
wavelength where the available $s$ range extends to cover several maxima/minima. This would be
possible at wavelength of 200--100~pm, which became available at LCLS recently and will be available
at upcoming facilities, \eg, the European XFEL, in the near future.

We do not observe direct signs of radiation damage in the diffraction data. While previous
experiments aimed specifically at the investigation of x-ray induced damage in strongly focused
x-ray beams~\cite{Young:Nature466:56, Erk:PRL110:053003}, here, we have actively avoided that regime
and performed the experiments using a hundred times larger cross-section of the x-ray beam. Under
these moderate-fluence conditions the damage can be rationalized based on simple cross-section
estimates for photoionization and elastic scattering and is detailed in the supplementary
information. Since the sample is replenished for every XFEL pulse, the diffractive imaging signal is
only sensitive to the dynamics of damaged molecules during the x-ray pulse ($\sim\!100$~fs). Using a
simple mechanical model we estimate that most ($\sim\!90$~\%) of the diffraction signal is due to
(practically) intact molecules. A minor fraction of the signal is due to damaged molecules with
small changes in molecular structure, which could not been resolved with the available x-ray
wavelength. Damage could even be mitigated using shorter ($\sim10$~fs) duration pulses; see
supplementary information for details. Moreover, an appropriate trade-off between pulse duration,
pulse energy, and repetition rate would allow the recording of atomically resolved x-ray diffraction
patterns of molecules within minutes~\cite{Barty:ARPC64:415}. At these high repetition rates one
could directly observe femtosecond molecular dynamics through snapshots for many time-delays in
pump-probe experiments of electronic-ground-state chemical dynamics.

In summary, we demonstrate the preparation of strongly aligned samples of polyatomic molecules at an
XFEL facility. We experimentally verify that the high-frequency, far off-resonant x-rays are an
ideal probe of alignment of molecular ensembles in an photoion momentum imaging approach. The
employed setup and conditions are applicable for coherent diffractive imaging of single biomolecules
or molecular ensembles. We show the possibility to perform spatially resolved single x-ray photon
counting. Due to the weak scattering signal from small isolated molecules, averaging of many shots
is necessary and possible for the observation of an analyzable diffraction signal, on top of a large
background from NIR photons. We confirm that the angular structures in the single molecule
diffraction patterns were preserved during averaging and that a diffraction pattern of isolated and
strongly aligned DIBN molecules was successfully measured beyond experimental noise. Even with the
experimentally limited range of scattering vectors $s$, the heavy-atom distance derived from the
$I(s)$-plot is in agreement with the computed molecular structure, demonstrating the capability to
extract structural information for small molecules.

Our results provide direct evidence for the feasibility of x-ray diffractive imaging of aligned
gas-phase ensembles of molecules. Analyzing radiation damage in detail shows that damage effects in
the diffraction pattern could be avoided by using shorter x-ray pulses with lower fluences at higher
repetition rates. This would allow to observe atomically resolved snapshots of ultrafast chemical
dynamics. Combined with advanced molecular beam delivery techniques, \eg, laser desorption or helium
droplet beams, considerably larger molecules could be delivered in cold beams, isomer selected, and
aligned, providing a bottom-up approach toward the envisioned atomic-resolution single-molecule
diffraction experiments. In contrast to ultrafast electron diffraction, pump-probe experiments with
x-ray pulses will not suffer from Coulomb-repulsion broadening or pump-probe velocity mismatch and
hence may permit better time resolution, \ie, in the range of 10-100 fs.

Parts of this research were carried out at the Linac Coherent Light Source (LCLS) at the SLAC
National Accelerator Laboratory. LCLS is an Office of Science User Facility operated for the
U.~S.~Department of Energy Office of Science by Stanford University. We acknowledge the Max Planck
Society for funding the development and operation of the CAMP instrument within the ASG at CFEL.
H.~S.\ acknowledges support from the Carlsberg Foundation. C.~C.\ and P.~J.\ acknowledge support
from the Swedish Research Council and the Swedish Foundation for Strategic Research. J.C.H.S.\ and
H.N.C.\ acknowledge NSF STC award 1231306. A.Ru.\ acknowledges support from the Chemical Sciences,
Geosciences, and Biosciences Division, Office of Basic Energy Sciences, Office of Science, US
Department of Energy. D.~R.\ acknowledges support from the Helmholtz Gemeinschaft through the Young
Investigator Program. This work has been supported by the excellence cluster ``The Hamburg Center
for Ultrafast Imaging -- Structure, Dynamics and Control of Matter at the Atomic Scale'' of the
Deutsche Forschungsgemeinschaft.

\clearpage

\begin{center}
\textbf{Supplementary Information:\\
  \titletext}%
\end{center}

\section{Analysis of the x-ray diffraction data}
\label{sec:analysis-x-ray}
\begin{figure}[t]
   \centering
   \includegraphics[width=\figwidth]{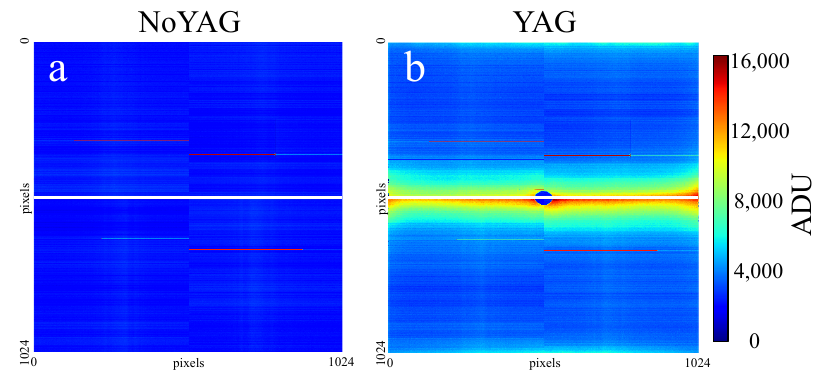}
   \caption{Experimentally obtained single shot raw diffraction patterns for not-aligned (\INoYAG,
      a) and aligned (\IYAG, b) DIBN molecules. All visible features are due to camera artifacts or
      laser background signals.}
   \label{fig:raw-patterns}
\end{figure}

\noindent%
Diffraction data was obtained for $\approx563\,000$ shots for aligned (YAG) and for
$\approx842\,000$ shots for unaligned (NoYAG) molecules, respectively. This corresponds to 7~h (YAG)
and 9~h (NoYAG) measurement time with LCLS operating at 60~Hz. \autoref{fig:raw-patterns} show
example single shot data for NoYAG (\INoYAG) (a) and YAG (\IYAG) (b) respectively. Different sources
of experimental background and pnCCD artifacts were taken into account in order to clean the single
shot data and extract single photon hits from the single shot data frames. The data contains a
pnCCD-channel dependent offset and gain variation, time-wise readout-variations (common mode), and
some channels (horizontally) or rows (vertically) with unusual high or low signal. The latter are
marked as ``bad pixel regions'' and were excluded from all further steps in data analysis. The pnCCD
detectors could not be fully shielded against NIR photons from the high-power YAG, resulting in
severe background signal levels and saturation of the lines closest to the central gap and at the
outer edges of the detector as can be seen in the single shot data in \autoref{fig:raw-patterns}\,b
for the YAG case. We extracted hitlists of x-ray photons after carefully subtracting all backgrounds
and data-acquisition-based artifacts from the pnCCD data frames. The saturated regions close to the
inner edge as well as hot pixels of the pnCCDs are neglected. The resulting data of x-ray hits
contain further real experimental background contained in these images, mostly due to scattering
from apertures, remaining helium seedgas, and rest gas (diffused helium, residual air) in the
chamber.

For a 2~keV photon the generated charge cloud created in silicon can spread over multiple pnCCD
pixels. This is considered in the analysis and such events are joined to single hits. In our
experiments, 64~\% of the recorded 2~keV photons are single-pixel hits, 35~\% are double pixel hits,
and $<1$~\% are spread over more pixels. Only single and double pixel hits were considered as
scattered x-rays in the analysis. A detailed account of this analysis will be published
elsewhere~\cite{Stern:LCLSanalysis:inprep}. Overall, this process results in lists of individual
photons and their position on the pnCCDs, with 0.20 x-ray photons per shot that are elastically
scattered from the molecular beam. These are histogrammed into diffraction patterns representing the
incoherent sum of the coherent x-ray-diffraction patterns of all molecules in the interaction
volume.

\begin{figure}[t]
   \centering
   \includegraphics[width=\figwidth]{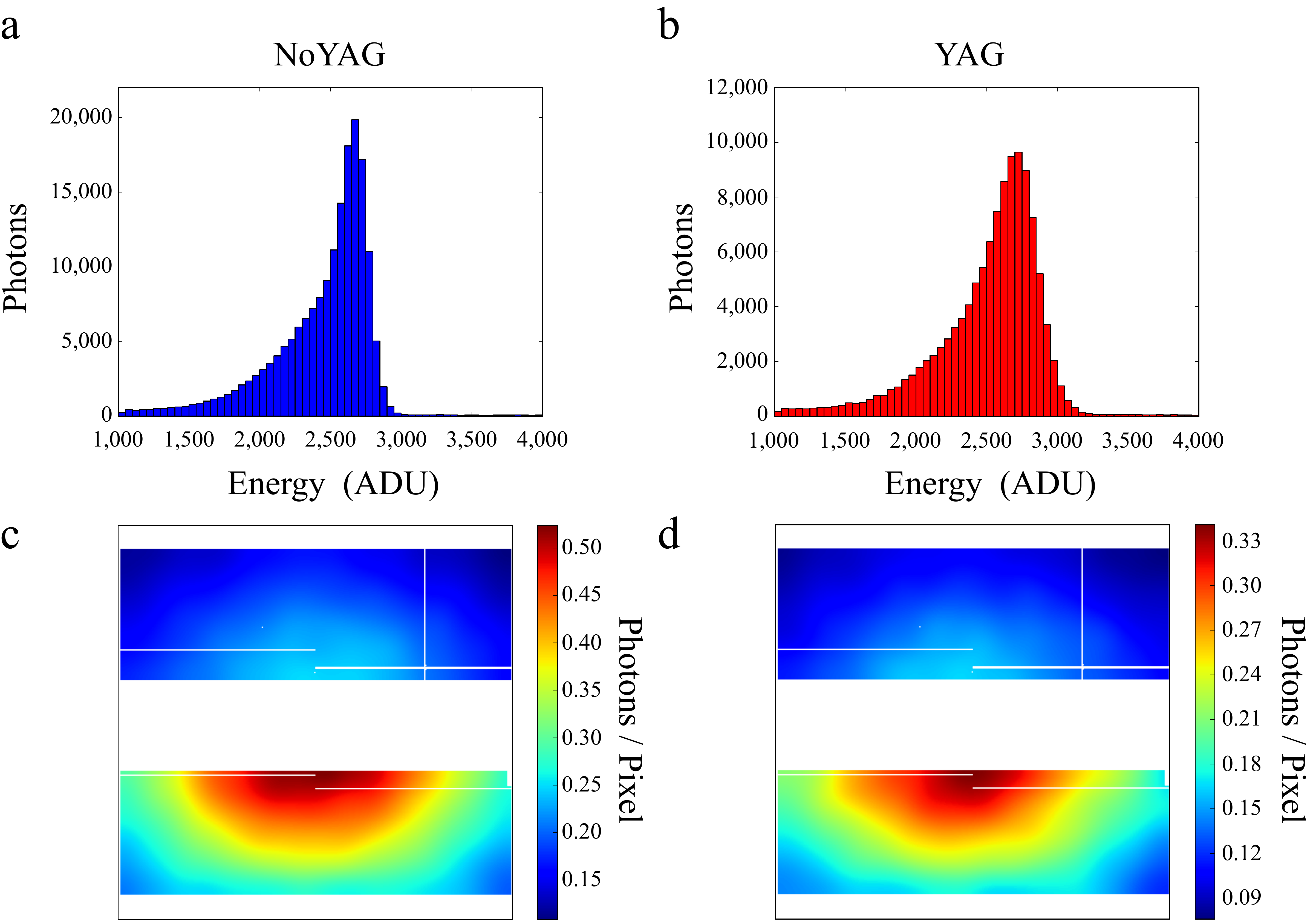}
   \caption{Spectra (a--b) and spatial distribution (c--d) of all photon hits from the completely
      cleaned data for the NoYAG and YAG case, neglecting ``hot-pixels'' and saturated regions in
      the center of the pnCCD detector.}
   \label{fig:corrected-patterns}
\end{figure}
In \autoref{fig:corrected-patterns} spectra of the completely cleaned data are shown for (a) \INoYAG
and (b) \IYAG respectively. 2~keV x-ray hits are expected at 2\,600~ADU. The width of the spectrum
is due to variations of the cleaned single shot frames after background subtraction, the energy
spread of the pnCCD detector, and the thresholding of the event recombination process when a single
photon hit is spread over more than a single pixel. The spatial distribution of all hits in the
energy interval 1\,500--3\,200~ADU is shown in \autoref{fig:corrected-patterns}~c--d. A difference
of the \INoYAG and \IYAG data is only weakly visible, even at strong degrees-of-alignment in our
experiment. In order to visualize the anisotropy, on top of a strong isotropic signal from the He,
the atomic scattering and experimental background, the isotropic part of the data was removed by
obtaining the diffraction-difference pattern. Therefore, after the NoYAG data (\INoYAG, \ie,
isotropically distributed DIBN) was scaled to the number of shots of the YAG data (\IYAG, \ie,
aligned DIBN), the NoYAG diffraction data was subtracted from the YAG data. The resulting
differencepattern is shown in Fig.~3~b in the main manuscript.

\section{Simulated x-ray-diffraction pattern}
\label{sec:simulated-pattern}\noindent%
Simulated diffraction patterns were calculated using an atomistic approach in which the target
molecule 2,5-diiodo\-benzo\-nitrile (C$_7$H$_3$I$_2$N, \DIBN) is considered to consist of single
atomic scatterers at fixed positions according to the calculated molecular structure. The molecular
scattering factor $F_{mol}(\mathbf{q})$ was derived by calculating the sum of all atomic scattering
factors $f_{j}(\mathbf{q})$ times their phase factors $e^{i\mathbf{qr_j}}$, hence
$F_{mol}(\mathbf{q})=\sum_{j} f_{j}(\mathbf{q})\cdot
e^{i\mathbf{qr}_j}$~\cite{Als-Nielsen:ModernXrayPhysics}. We used atomic scattering factors from
ref.~\citealp{Waasmaier:ActaCrystA51:416} and dispersion corrections to the scattering factors from
ref.~\citealp{Henke:ADNDT54:181}. For a given set of parameters, \ie, photon energy,
number-of-photons per XFEL-pulse, number-of-shots, number density \M~of target molecules in our
molecular beam and the given geometry of the detector, the expected number of x-ray photons
scattered to a certain detector area was calculated. The experimental diffraction pattern of an
ensemble will be blurred with respect to the diffraction pattern from a single molecule due to the
finite (\ie, non-perfect) alignment. The diffracted intensity from a non-perfectly aligned molecular
ensemble is obtained by the convolution of diffraction patterns of possible orientations distributed
and an alignment angular distribution $n(\theta)$. The latter is given by
$n(\theta)=\exp(-\sin^2\theta/(2\sigma^2))$~\cite{Friedrich:PRL74:4623}, where $\theta$ is the angle
between the axis given by the (linear) YAG polarisation and the molecular axis of highest
polarizability, which is approximately, within a few degree, the same as the I--I axis. In order to
analyse the experimentally obtained diffraction pattern, diffraction of aligned and not- aligned
DIBN with a strong background from the He seed gas was simulated. Furthermore, for fitting the main
structural feature, \ie, the I--I distance, to the measured diffraction data (Fig.~4 in main text),
the iodine-iodine distance was varied in the range of 500--1000~pm.

\begin{figure}
    \centering
    \includegraphics[width=\linewidth]{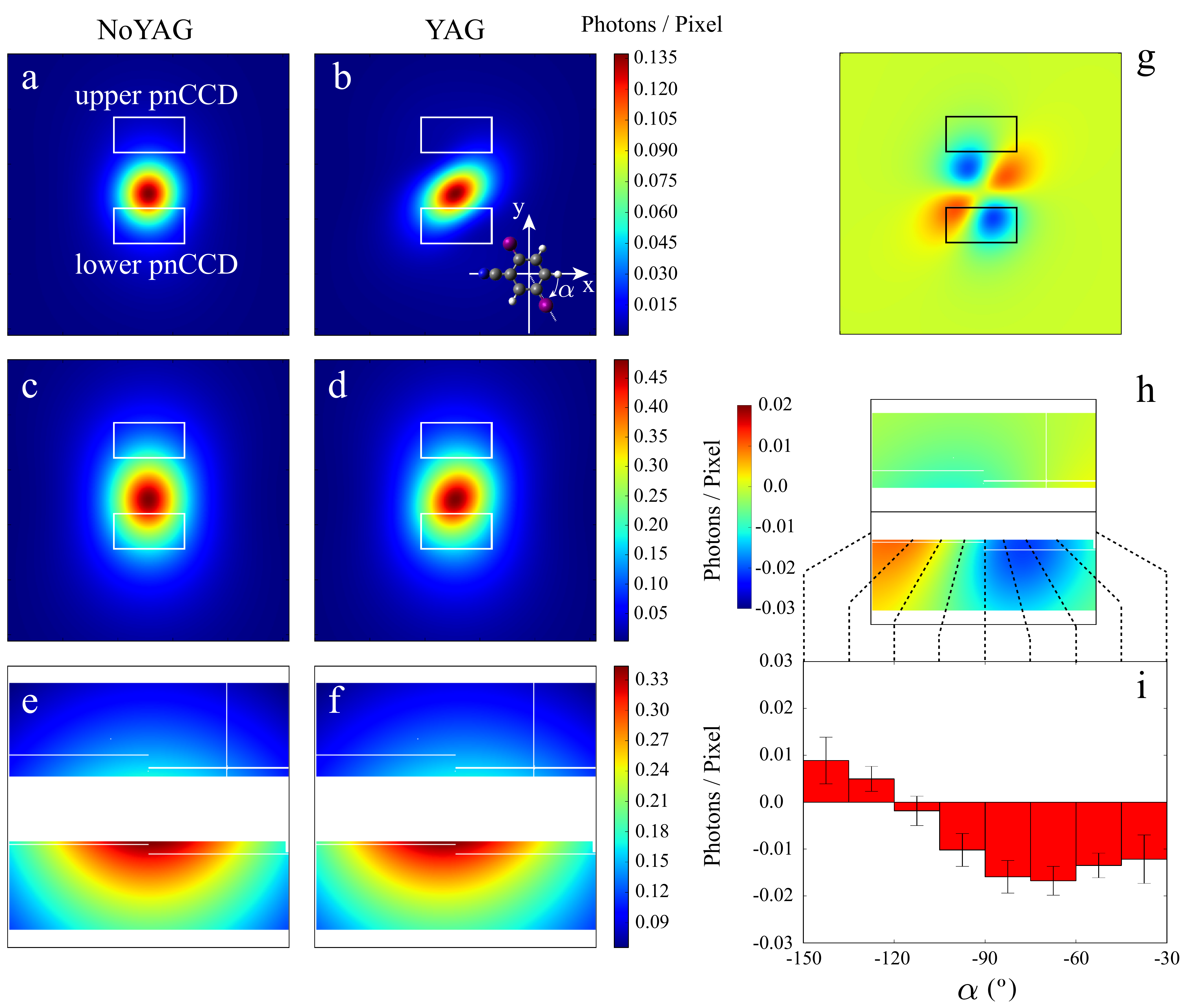}
    \caption{Simulated data of DIBN for not-aligned (\INoYAG) and aligned molecules (\IYAG,
       polarization $\alpha=-60\degree$) with FEL parameters as in the experiment. (a,b) show
       diffraction of DIBN only, (c,d) show diffraction of DIBN with a strong scattering background
       of He added to the diffraction from DIBN (He/DIBN=14,000). The white rectangles mark the
       position of the pnCCD panels in the experiment. (e,f) show the simulated intensities in the
       NoYAG and YAG case in the regions covered by the pnCCD panels. The diffraction-difference
       (\IYAGmNoYAG) pattern is shown in (g--h), along with the azimuthal histogramm visualizing the
       angular anisotropy (i). See text for details.}
    \label{fig:Simulation}
\end{figure}
\autoref{fig:Simulation} shows simulated data for the parameters of the experiment (563,453 shots as
in the YAG case, 4.375$\cdot$10$^{12}$ incident photons/shot). The molecular beam density of DIBN is
$\M = 10^{8}$~cm$^{-3}$. The degree-of-alignment of DIBN in the YAG case was $\cost = 0.83$ with
respect to the axis of the YAG polarisation of $\alpha = -60\degree$. The first row (a,~b) shows the
scattering intensity (in terms of photons/pixel) for DIBN only. White rectangular frames mark the
position of the top and bottom pnCCD panel in the experiment. The second row (c,~d) shows the
scattering intensity of DIBN (like in a,~b), but with a strong background of 14\,000 He atoms added
per DIBN. The amount of He was chosen to match the total measured intensity (with keeping the
molecular beam density \M~of DIBN at $\M=10^{8}$~cm$^{-3}$). It should be noted that in the
experiment the amount of He is lower due to strong contributions from other sources of background,
such as rest-gas scattering, stray light from apertures, etc., which here is effectively simulated
by the He signal. Unfortunately, the experimental background without molecular beam was not measured
sufficiently long, but the exact amounts of He and other background sources become obsolete when the
data is analyzed in terms of the diffraction-difference, \ie, when taking the difference image of
aligned and isotropically-distributed DIBN (\IYAGmNoYAG).

In the lower row (e,~f), the simulated intensities in the NoYAG and YAG case are shown for the
regions covered by the pnCCD panels in the experiment. The alignment $\cost~=~0.83$ of DIBN results
in a slight shift of the 0-order scattering maximum towards the left with respect to the center, but
the effect is hardly visible due to the background. The difference (and hence the anisotropy in the
YAG case) can be visualized best by plotting the difference of both diffraction patterns as shown in
(g--i).

\section{Ionization and radiation damage}
\label{sec:damage}

Ionization and two-dimensional momentum distributions of \Iplus\ ions were measured to quantify the
degree of alignment as shown in Fig.~2 in the main manuscript. When utilizing the FEL instead of the
TSL, the two \Iplus\ peaks (d) were further separated than the ones in (b), with the latter
corresponding to \Iplus\ ions from doubly and triply ionized molecules~\cite{Larsen:JCP111:7774}.
Thus, the DIBN molecules accumulate overall higher charges when ionized by the x-rays, indicating
molecular radiation damage processes and the possibility to observe these processes in ion-imaging
experiments~\cite{Erk:PRL110:053003}. In time-of-flight measurements we observed iodine ions in
charge states up to I$^{+7}$; I$^{+8}$ cannot be observed because it is hidden by the much larger
O$^+$ signal from residual air. With a 2~keV photon creating an M-shell vacancy in iodine, one
expects four or more charges per photon~\cite{Kochur:JPB27:1709}, which are quickly redistributed
over the molecule~\cite{Erk:PRL110:053003}. From the observed monotonically decaying intensities of
the I$^{+n}$ ion signals, we conclude that typically one and, generally, at most two photons are
absorbed per molecule. Calculation of the probability $p_\text{abs}$ of a single iodine absorbing a
2~keV photon and subsequently getting ionized during a single FEL shot is based on the
photoabsorption cross section of iodine $\sigma_{\text{abs}} =
0.4192$~Mbarn~\cite{Berger:XCOM1.5:2010}. The probability is given by $p_{\text{abs}} =
\sigma_{\text{abs}}\cdot{}N_{\text{photons}}/A_0$ with the number of photons $N_{\text{photons}}$
impinging on the interaction area $A_0$. With the FEL-pulse energy in the interaction region of
1.4~mJ, the number of 2~keV photons is $N_{\text{photons}} = 4.375\cdot10^{12}$. The interaction
area ($\omega=30$~$\mu$m) is $A_0 = 7.068\cdot 10^{-6}$~cm$^2$. Hence, the probability for a single
iodine atom to absorb a 2~keV photon from the FEL pulse photons is $p_\text{abs} = 0.25$ and since
the a DIBN molecule is made of 2 iodines, the probability for a DIBN molecule to absorb a 2~keV
photon is 0.5. Thus, half of the DIBN molecules absorb a 2~keV photon during the FEL pulse and will
eventually be multiply ionized by Auger relaxation and fragmented due to Coulomb explosion. The
diffraction pattern of fragmenting DIBN will look different than the diffraction from intact DIBN.

Here, only nuclear damage is considered. In the following, the influence of scattering from
fragmenting DIBN on the diffraction pattern is discussed in terms of an effective spatial
distribution of the two main scattering centers (\ie, the two iodine atoms) ``seen'' by the FEL. The
velocity distribution of \Iplus~ions upon ionization by the FEL is obtained from the measured
momentum distribution shown in Fig.~2 of the main text. This distribution can be well approximated
by a Gaussian with mean $\mu=2\,700$~m/s and $\sigma=700$~m/s. Since one is interested in the
relative velocity of the two iodines (rather than in the velocity of \Iplus~in the lab frame) it is
assumed that DIBN fragments into \Iplus~and C$_7$H$_3$IN$^{\text{n}+}$. Momentum conservation yields
a velocity distribution $v_{\text{ion}}$ of the relative velocity of \Iplus~with respect to
\ce{C7H3IN^{+n}} with $\mu=4200$~m/s and \mbox{$\sigma=1090$~m/s}. The following discussion is
restricted to this simple case since a complete velocity distribution of all ions has not been
determined experimentally.

\begin{figure}[t]
    \centering
    \includegraphics[width=0.5\linewidth]{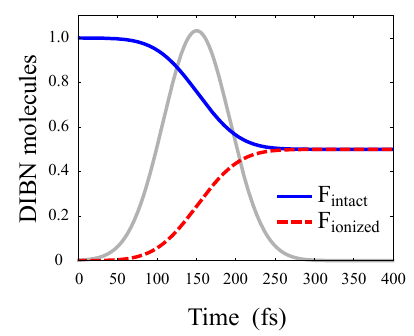}
    \caption{The fractions of intact and ionized DIBN as a function of time for a 100~fs (FWHM) FEL
       pulse indicated by the grey line.}
    \label{fig:FELpulse}
\end{figure}
The velocity distribution is utilized in order to estimate an effective spatial distribution of the
two iodine atoms in DIBN as seen by the FEL pulse. The pulse duration was 100~fs, estimated from the
electron bunch length and pulse duration measurements~\cite{Duesterer:njp13:093024}. In
\autoref{fig:FELpulse} the fractions of intact and ionized DIBN over the duration of the x-ray pulse
is plotted. At each time $t_i$ the effective I--I distribution seen by the incident x-ray intensity
$I_{\text{FEL}}(t_i)$ is calculated. The effective I--I distribution $s(t_i)$ depends on all times
$t_j<t_i$ at which molecules were ionized and started to recoil with the given velocity
distribution:

\begin{align}
   s(t_i) = \; & I_{\text{FEL}}(t_i) \notag\\
   & \cdot \left[\int\limits_{0}^{t_i} \mathit{v}_{\text{ion}}\cdot(t_i-t_j)\cdot f_\text{ionized}(t_j)\;dt_j+F_\text{intact}(t_j)\right]
   \label{eq:EffSpatialDistr}
\end{align}
where $F_{\text{ionized}}$ and $F_{\text{intact}}$ are the fractions of molecules that are ionized
and intact at time $t_i$, respectively. The amount of ionized molecules at $t_i$ depends on all
times $t_j$ before, hence $F(t_i) = \int_{0}^{t_i} f_{\text{ionized}}(t_j)$. The onset of Coulomb
explosion is assumed to happen instantly upon absorption of a 2~keV photon, neglecting the finite
delay due to Auger decay and subsequent charge reorganisation within the DIBN molecule. This is a
worst case approximation which rather overestimates the distance of the fragmenting ions. The
complete effective spatial distribution $S$ seen by the whole FEL pulse is then just the sum over
all $t_i$
\begin{equation}
   S=\int s(t_i)dt
   \label{eq:EffSpatialDistrTotal}
\end{equation}

\begin{figure}[t]
    \centering
    \includegraphics[width=\linewidth]{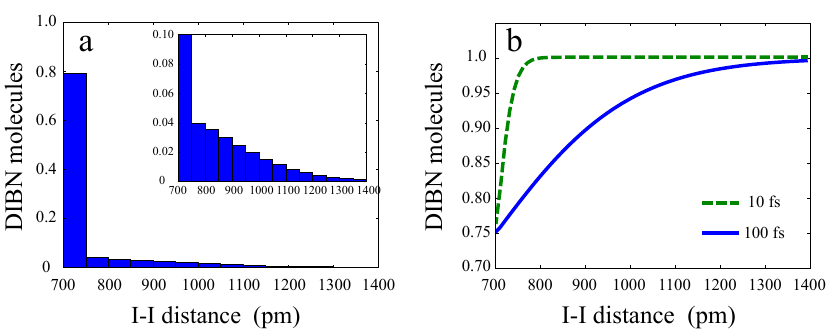}
    \caption{(a) Histogram of $S$, visualizing the fraction of molecules in different distance
       intervals, seen by the FEL pulse. (b) The cumulative distribution of $S$ (for smaller
       stepsize in I-I-distance).}
    \label{fig:IodineDistanceDistr}
\end{figure}
\autoref{fig:IodineDistanceDistr}~a shows a histogram of $S$ to visualize the fraction of molecules
in different distance intervals as ``seen'' by the FEL pulse. The cumulative distribution is given
in \autoref{fig:IodineDistanceDistr}~b. For a FEL pulse duration of 100~fs (solid-blue) this shows
that 75~\% of the elastically scattered photons originate from scattering at intact molecules and
that another 15~\% (20~\%) of the scattered photons originate from scattering of molecules
corresponding to I--I distances that are less than 200~pm (330~pm) longer than the
700~pm-equilibrium distance. These changes are not significant in the current experiment, \ie, while
the resolution using 620~pm wavelength radiation (and the limited range of $s$-values) is not high
enough to analyze these changes, the experimentally observed elongated I--I distance could be due to
these damage effects.

If shorter x-ray pulses, \eg, of $\sim10$~fs duration, would be used (dashed-green line in \autoref
{fig:IodineDistanceDistr}~b) practically no damage would be observed even for the same pulse energy.
Thus, for the diffraction from ensembles of isolated ``small'' molecules even the dose delivered, in
a single pulse, by a state-of-the art very intense x-ray laser is small enough to record diffraction
patterns without radiation damage. For molecules with larger photoabsorption cross sections, the
pulse energy could always be reduced to recover this regime, at correspondingly extended averaging
times. With upcoming high-repetition rate light sources, for instance the European XFEL in Hamburg,
Germany, this would allow the recording of atomically resolved x-ray diffraction patterns of
molecules within minutes~\cite{Barty:ARPC64:415}. Moreover, at these high repetition rates one could
directly observe femtosecond molecular dynamics through snapshots for many time-delays in pump-
probe experiments of electronic-ground-state chemical dynamics.

\end{document}